\documentclass[floats,preprint,prl,aps]{revtex4}
\usepackage{latexsym}
\usepackage{graphicx}
\newcommand\beq{\begin{equation}}
\newcommand\eeq{\end{equation}}
\newcommand\bea{\begin{eqnarray}}
\newcommand\eea{\end{eqnarray}}
\newcommand\non{\nonumber}
\newcommand\bib{\bibitem}
\begin{document}
%\textheight=24cm
%\twocolumn[\hsize\textwidth\columnwidth\hsize\csname@twocolumnfalse\endcsname
\title{\bf Quantum criticality of geometric phase in coupled  
optical cavity arrays under linear quench}

\author{\bf Sujit Sarkar}
\address{\it Poornaprajna Institute of Scientific Research,
4 Sadashivanagar, Bangalore 5600 80, India.\\
e-mail: sujit.tifr@gmail.com\\
phone: 0091-80-23611836\\
Fax: 0091-80-23619034 \\
}
\date{\today}

\begin{abstract}
The atoms trapped in microcavities and interacting through the exchange of virtual
photons can be modeled as an anisotropic Heisenberg spin-1/2 lattice.
We study the dynamics of the geometric
phase of this system under the linear quenching process of laser field detuning 
which shows the XX criticality of the geometric phase 
in presence of single Rabi frequency oscillation. We also study the quantum
criticality for different quenching rate 
in the presence of single or two Rabi frequencies oscillation
in the system. 
\\
PACS: 42.50.Pq, 03.65.Vf, 42.50.-p\\
Keywords: Cavity Quantum Electrodynamics, Geometric Phase and Quantum Optics.\\
% 42.50.Pq, Cavity Quantum Electrodynamics\\
% 03.65.Vf, Phases: Geometric, Dynamic or Topological \\
% 42.50.-p, Quantum Optics\\ 
\end{abstract}
%\vskip .5 true cm
\maketitle

%\section{I. Introduction}
%\vskip -1.5 true cm
{\bf Introduction:}  
The recent experimental success in engineering strong interaction 
between the photons and atoms in high quality micro-cavities opens up
the possibility to use light matter system as quantum simulators for
many body physics [1-16]. 
%\cite{green,hart1,hart2,ji,byrn,caru,bhas,toma,zhao,pipp,brij,horn,blat,
%bose,hur1,sujop,horo}.
The authors of Ref. (\cite{hart1}),(\cite{hart2}) and (\cite{sujop}) have shown that
effective spin lattice can be generated with individual atom in the
micro-cavities that are coupled to each other via exchange of virtual
photons. The two states of spin polarization are represented by the
two long lived atomic levels in the system.\\
A Many body Hamiltonians can be created and probed in coupled cavity arrays. 
In our previous study, we have explained explicitly the basic physics
of the formation of micro-optical cavity \cite{sujop}.
The atoms in the cavity are used for detection and also for generation of
interaction between photons in the same cavity.
As the distance between the adjacent cavities is considerably
larger than the optical wave length of the resonant mode, individual cavities
can be addressed. This artificial system can act as a quantum simulator.
In this optical cavities system one can find 
different quantum phases of polariton ( a combined excitations of atom-photon
interactions. ) by using the spin model
that conserve the total number of excitations.
This micro-cavity system shows the different quantum phases and 
quantum phase transitions.\\ 
Quantum Phase Transition (QPT) associate with the fundamental changes that occurs
in the macroscopic nature of the matter at zero temperature due to
the variation of some external parameter. The quantum phase transitions
are characterized by the drastic change in the ground state
properties of the system driven by the quantum fluctuations \cite{ss}. \\
Motivation of this research paper:\\
To the best of our knowledge, 
here we not only study the dynamics of geometric phase but also 
solve the nature of criticality
explicitly  
under laser field detuning quenching process for different quenching
rate which is absent in the previous literature of cavity QED [1-16].\\ 
Here we mention very briefly the essence of the geometric phase in
the condensed matter.  
The geometric phases have been associated with a
variety of condensed matter phenomena
\cite{thou,resta,hatsugai,bb} since its inception
\cite{berry}. Besides, various theoretical investigations, the geometric
phases have been experimentally tested in various cases, e.g. with
photons \cite{p1,p2,p3}, with neutrons \cite{n1,n2} and with atoms
\cite{a1}. 
The quantum state engineering of cavity QED is in the state of art
due to the rapid experimental/technological progress of the
subject. We hope that the theoretical scheme which we propose
for the laser field detuning induce quenching process in 
the dynamics of geometric phase will be predicted by
the experimental group based on photon.\\ 
The generation of the geometric phase (GP) is a witness of
a singular point in the energy spectrum that arises in all
non-trivial geometric evolutions. In this respect, the connection of
the geometric phase with quantum phase transition (QPT) has been
explored very recently \cite{car,zhu,hamma,sujgeo}. The geometric phase can
be used as a tool to probe QPT in many body systems. Since response
times typically diverge in the vicinity of the critical point,
sweeping through the phase transition with a finite velocity leads
to a breakdown of adiabatic condition and generate interesting
dynamical (non-equilibrium ) effects. In the case of thermal phase
transitions, the Kibble-Zurek (KZ) mechanism \cite{kib,zur} explains
the formation of defects via rapid cooling. This idea of  defect
formation in second order phase transition has been extended to zero
temperature QPT \cite{zur1,dziar} by
studying the spin models under
linear quench. We will use this concept in the present study.\\
The micro-cavities of a photonic crystal are coupled through the exchange of photons.
Each cavity consists of one atom with three levels in the energy
spectrum, two of them are long lived
and represent two spin states of the system and the other represent excited
states Ref. \cite{hart1,hart2,sujop}). Externally
applied laser and cavity modes couple to each atom of the cavity. It may 
induce the Raman transition between these two long lived energy levels. Under a
suitable detuning between the laser and the cavity modes, virtual photons
are created  
in the cavity which mediate interactions with another atom in a 
neighboring cavity. One can eliminate the excited states 
by choosing the appropriate detuning between the applied laser
and cavity modes. 
Then one can achieve only two states per atom in the long 
lived state and the system can be described by a spin-1/2 
Hamiltonian \cite{hart1,hart2,sujop}.\\ 

The Hamiltonian of our present study consists of three parts:
\beq
H ~= ~ {H_A} ~+~ {H_C}~+~{H_{AC}}
\eeq     
Hamiltonians are the following
\beq
{H_A} ~=~ \sum_{j=1}^{N} { {\omega}_e } |e_j > <e_j | ~+~ 
{\omega}_{ab} |b_j > <b_j | 
\eeq
where $j$ is the cavity index. ${\omega}_{ab} $ and ${\omega}_{e} $ are 
the energies of the state $ | b> $ and the excited state respectively. The
energy level of state $ |a > $ is set as zero. $|a>$ and $|b> $ are
the two stable state of a atom in the cavity and $|e> $ is the
excited state of that atom in the same cavity. 
The following Hamiltonian describes the photons in the cavity,
\beq
 {H_C} ~=~ {{\omega}_C} \sum_{j=1}^{N} {{c_j}}^{\dagger} {c_j} ~+~
{J_C} \sum_{j=1}^{N} ({{c_j}}^{\dagger} {c_{j+1}} + h.c ),  
\eeq 
where ${c_j}^{\dagger}({c_j})$ is the photon
creation(annihilation) operator for the photon field in the $j $'th cavity, ${\omega}_C $
is the energy of photons and $ J_C $ is the tunneling rate of photons
between neighboring cavities.
The interaction between the atoms and photons and also by the driving lasers
are described by
\beq
{H_{AC}}~=~ \sum_{j=1}^{N} [ (\frac{{\Omega}_a}{2} e^{-i {{\omega}_a} t} +
{g_a} {a_j}) |e_j > < a_j | + h.c] + [a \leftrightarrow b ] . 
\eeq
Here ${g_a} $ and ${g_b} $ are the couplings of the cavity mode for the
transition from the energy states $ |a > $ and $ | b> $ to the excited state.
${\Omega}_a $ and ${\Omega}_b $ are the Rabi frequencies of the lasers
with frequencies ${\omega}_a $ and $ {\omega}_b $ respectively.\\
The authors of Ref. \cite{hart1,hart2,sujop}
have derived an effective spin model by considering the following physical
processes:
A virtual process regarding emission and absorption of
photons between the two stable  states of neighboring cavity yields the resulting 
effective Hamiltonian as
\beq
{H_{xy}} = \sum_{j=1}^{N}  B {{\sigma}_j}^{z} ~+~\sum_{j=1}^{N} 
(\frac{J_1}{2} {{\sigma}_j}^{\dagger} {{\sigma}_{j+1}}^{-} ~+~
\frac{J_2}{2} {{\sigma}_j}^{-} {{\sigma}_{j+1}}^{-} + h.c )
\eeq 
When $J_2 $ is real then this Hamiltonian reduces to the XY model.
Where ${{\sigma}_j}^{z} = |b_j > <b_j | ~-~ |a_j > <a_j | $,
${{\sigma}_j}^{+} = |b_j > <a_j | $, ${{\sigma}_j}^{-} = |a_j > <b_j | $
. 
%\beq
%H_{xy}~=~ \sum_{i=1}^{N} B ( {{\sigma}_i}^{z}~+~{J_x} {{\sigma}_i}^{x}
%{{\sigma}_{i+1}}^{x} ~+~ {J_y} {{\sigma}_i}^{y}
%{{\sigma}_{i+1}}^{y}) .
%\eeq
\bea
H_{xy} & = & \sum_{i=1}^{N} ( B  {{\sigma}_i}^{z}~+~ {J_1} ( {{\sigma}_i}^{x}
{{\sigma}_{i+1} }^{x}  + {{\sigma}_i}^{y}
{{\sigma}_{i+1} }^{y} ) \non\\
& & + {J_2} ( {{\sigma}_i}^{x}
{{\sigma}_{i+1} }^{x} - {{\sigma}_i}^{y}
{{\sigma}_{i+1} }^{y} ) ) \non\\    
& & =  \sum_{i=1}^{N} B ( {{\sigma}_i}^{z}~+~{J_x} {{\sigma}_i}^{x}
{{\sigma}_{i+1}}^{x} ~+~ {J_y} {{\sigma}_i}^{y}
{{\sigma}_{i+1}}^{y}) .
\eea
With ${J_x} = (J_1 + J_2 ) $ and ${J_y} = (J_1 - J_2 ) $.\\
We follow the references \cite{jame,hart1} to present the analytical 
expression for the different physical parameters of the system.\\
\beq
B = \frac{\delta_1}{2} - \beta 
\eeq
\bea
\beta  & = &  \frac{1}{2} [\frac{{|{\Omega_b}|}^2 }{4 {\Delta}_b }
({\Delta}_b - \frac{{|{\Omega_b}|}^2 }{4 {\Delta}_b } - \non\\ 
& & \frac{{|{\Omega_b}|}^2 }{4 ( {\Delta}_a  - {\Delta}_b )} - {\gamma_b} {g_b}^2 
- {\gamma_1} {g_a}^2 + {\gamma_1}^2 \frac{{g_a}^4 }{{\Delta_b}} - (a \leftrightarrow b)] 
\eea
\beq
{J_1} = \frac{\gamma_2}{4} ( \frac{{|{\Omega_a}|}^2 {g_b}^2 }{{ {\Delta}_a }^2 }
 +  \frac{{|{\Omega_b}|}^2 {g_a}^2 }{{ {\Delta}_b }^2 } )
\eeq
\beq
{J_2} = \frac{\gamma_2}{2} ( \frac{{\Omega_a} {\Omega_b} g_a g_b }{{\Delta}_a {\Delta_b} }
 ).
\eeq
Where 
$ \gamma_{a,b} = \frac{1}{N} \sum_{k} \frac{1}{ {\omega}_{a,b} - {\omega}_k } $
$ \gamma_{1} = \frac{1}{N} \sum_{k} \frac{1}{ ( {\omega}_{a}+  {\omega}_{b})/2 - {\omega}_k } $ and
$ \gamma_{2} = \frac{1}{N} \sum_{k} \frac{e^{ik} }{ ( {\omega}_{a}+  {\omega}_{b})/2 - {\omega}_k } $
${\delta_1} = {\omega}_{ab} - ({\omega}_a - {\omega}_b )/2 $, 
${\Delta}_a = {\omega}_e - {\omega}_a$.  
${\Delta}_b = {\omega}_e - {\omega}_a -({\omega}_{ab} - {\delta_1})$.
${{\delta}_a}^{k} = {\omega}_e - {\omega}_k $,
${{\delta}_b}^{k} = {\omega}_e - {\omega}_k  -({\omega}_{ab} - {\delta_1}) $,
$g_a$ and $g_b$ are the couplings of respective transition to the cavity mode,
${\Omega}_a $ and ${\Omega}_b$ are the Rabi frequency of laser with frequency
$\omega_a $ and $\omega_b $. \\

%%%%%%%%%%%%%%%%%%%%%%%%%%%%%%%%%%%%%%%%%%%%%%%%%%%%%%%%%%%%%%%%%%%%%%%%%%%%%%%%%% 
{\bf Model Hamiltonian and Quantum Phases:}
We express our model Hamiltonian in a more explicit way as
\beq
H ~=~ \sum_n ~[  (1+ \alpha) ~S_n^x S_{n+1}^x ~+~ (1-\alpha)~ S_n^y S_{n+1}^y
+ B \sum_n ~S_n^z  
\label{ham2}
\eeq
where $S_n^{\alpha}$ are the spin-1/2 operators.
We assume that the $XY$ anisotropy  
$ 0 < \alpha \leq 1 $
and magnetic field strength is $h \ge 0$.
The parameters correspondence
between the micro cavities and spin chain are the following:
${J_1 =1 }$ and ${J_2} =\alpha $. 
\begin{figure}
\includegraphics[scale=0.50,angle=0]{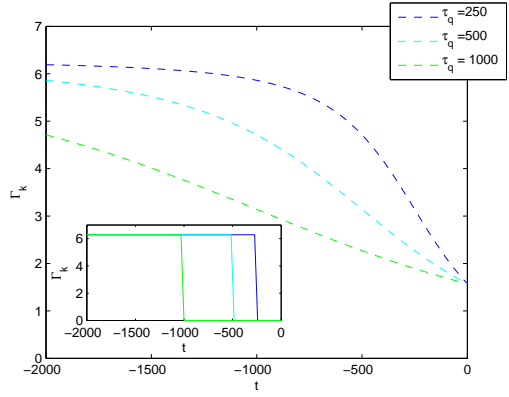}
\caption{Color online, variation of geometric phase with time
for different quenching time for $k =\pi/3 $. The anisotropy parameter ($\alpha =1$, transverse Ising model)
for the dash curves and inset for $\alpha =0 $ (isotropic XX spin model).} 
\label{Fig. 1 }
\end{figure}

\begin{figure}
\includegraphics[scale=0.50,angle=0]{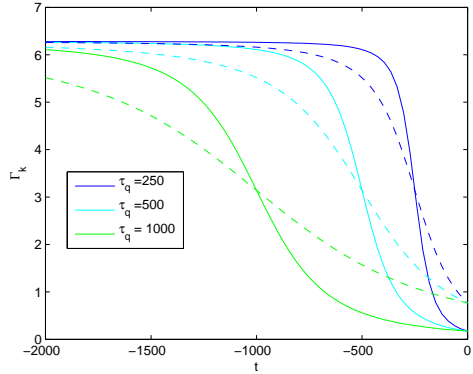}
\caption{Color online, variation of the geometric phase
with time for three different quenching time for $k =\pi/3 $. The solid curve is
for the $\alpha =0.2$ and dashed curve is for $\alpha =0.8 $.
 }
\label{Fig. 2 }
\end{figure}

\begin{figure}
\includegraphics[scale=0.50,angle=0]{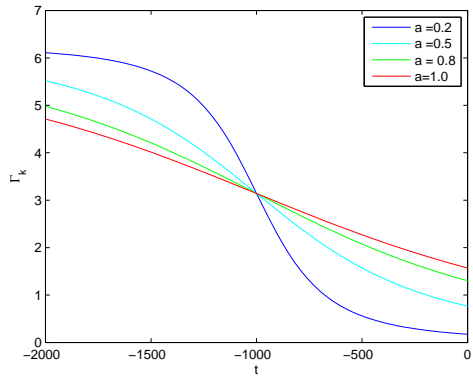}
\caption{Color online, variation of geometric phase with time
for the quenching time ${\tau_q = 1000}$ for different
anisotropy in exchange couplings for $k =\pi/3$. The all curves are meeting
in the point at $ t= {\tau_q }$. } 
\label{Fig. 3 }
\end{figure}
%%%%%%%%%%%%%%%%%%%%%%%%%%%%%%%%%%%%%%%%%%%%%%%%%%%%%%%%%%%%%%%%%%%%%
{\it Geometric Phase and Criticality}:
Here, we calculate the geometric phase and its dynamics
under the quenching of magnetic field.
In this model, the geometric phase of the ground state is evaluated
by applying a rotation of $\phi$ around the
$z$-axis in a closed circuit to each spin \cite{car, car1}.
A new set of Hamiltonians $H_\phi$ is constructed from the Hamiltonian
 (H) as
 \begin{equation}\label{h2}
 H_\phi=U(\phi)~H~U^\dagger(\phi)
 \end{equation}
 where
% \begin{equation}\label{g}
$U(\phi)=\prod_{j=-M}^{+M} \exp(i\phi\sigma_j^z/2) $
% \end{equation}
 and $\sigma_j^z$ is the $z$ component of the standard Pauli matrix at site $j$.
Here $M$ is the integer which relates with the lattice site numbers by the
following relation $ 2 M + 1 = N $. In our atom-cavity system, the rotation of
laser field around the z-axis is equivalent to the rotation of the quantum
spin system around the z-axis. 
The family of Hamiltonians generated by varying $\phi$ has the same energy spectrum as
the initial Hamiltonian and $H(\phi)$ is $\pi$-periodic in $\phi$.
The ground state  $|g> $  of the system is expressed as
 \begin{equation}\label{g}
|g> =\prod_{k>0}(\cos\frac{\theta_k}{2}|0>_k |0>_{-k}
-i\exp(2i\phi)\sin\frac{\theta_k}{2}|1>_k|1>_{-k} )
 \end{equation}
 where $|0>_k$ and $|1>_k$ are the vacuum and single fermionic excitation of the
$k$-th momentum mode respectively. The angle $\theta_k$ is given by
\begin{equation}
 %\cos\theta_k=\frac{\cos\frac{2\pi k}{N}-B}{\Lambda_k},
\cos\theta_k=\frac{\cos k-B}{\Lambda_k}
 \end{equation}
 and
% \begin{equation}
%\displaystyle{ \Lambda_k=\sqrt{(\cos \frac{2\pi k}{N}-B)^2+\gamma^2\sin^2\frac{2 \pi k}{N}}}%
$ \displaystyle{ \Lambda_k=\sqrt{(\cos k-B)^2+\alpha^2\sin^2 k}} $
% \end{equation}
is the energy gap above the ground state.
The ground state is a direct product of $N$ spins, each lying in the
two-dimensional Hilbert space spanned by $|0>_k
 |0>_{-k}$ and $|1>_k |1>_{-k}$. For each value of $k$,
the state in each of the two dimensional Hilbert space can be represented
as a Bloch vector with coordinates $(2\phi,\theta_k)$.
In our previous study \cite{sujop,bonodi}
we have shown explicitly the derivation
of geometric phase for this problem.\\
 \begin{equation}\label{ph}
  \Gamma_k=\pi(1-\cos\theta_k)
\end{equation}
 The total geometric phase of the state $|g>$ is given by
% \begin{equation}
%\Gamma_g=\sum_{k >0}~\pi(1-\cos\theta_k)
$ \Gamma_g=\sum_{k}~ \Gamma_k $
% \end{equation}
. For an adiabatic evolution, if the initial state is an eigenstate,
the evolved state remains in the eigenstate.  
Now we derive the instantaneous geometric phases of this system
due to a gradually decreasing magnetic field, i.e., the quenching field.
\begin{equation}\label{quench}
 B(t<0)=-\frac{t}{\tau_q}
 \end{equation}
$B(t)$, driving the transition, is assumed to be linear with an
adjustable time parameter $\tau_q$ ($1/{\tau_q} $ is the quenching
rate).
The system be initially at time $ t(<0)<<\tau_q$ such that $B(t)>>1$,
in our parameter space of cavity arrays system,
$ {\delta_1} (t) >> [\frac{{|{\Omega_b}|}^2 }{4 {\Delta}_b }
({\Delta}_b - \frac{{|{\Omega_b}|}^2 }{4 {\Delta}_b } -  
 \frac{{|{\Omega_b}|}^2 }{4 ( {\Delta}_a  - {\Delta}_b )} - {\gamma_b} {g_b}^2 
- {\gamma_1} {g_a}^2 + {\gamma_1}^2 \frac{{g_a}^4 }{{\Delta_b}} - (a \leftrightarrow b)] $.
Here we consider the quenching of the magnetic field by considering quenching the detuning
field ${\delta}_1 $.\\  
The instantaneous ground state at any instant $t$
is given by
\begin{equation}\label{g}
 |\psi_0(t)>=\prod_{k}(\cos\frac{\theta_k(t)}{2}|0>_k|0>_{-k}
-i\exp(2i\phi)\sin\frac{\theta_k(t)}{2}|1>_k|1>_{-k} )
\end{equation}
%where the angle $\theta_k(t)$ is then given by
%\begin{equation}
 %\cos\theta_k(t)=\frac{\cos k -(-\frac{t}{\tau_q})}{\sqrt{(\cos k -\frac{-t}{\tau_q})^2+\gamma^2\sin^2 k}},
 %\end{equation}
We now use eqn. (14) and (13) to derive
 the geometric phase of the $k^{th}$ mode  which yields
 \begin{equation}
\Gamma_k(t)=\pi\left(1-\frac{\cos k +\frac{t}{\tau_q}}
{\sqrt{(\cos k +\frac{t}{\tau_q})^2+\alpha^2\sin^2 k}}\right)
\end{equation}
The geometric phase for an isotropic system with $\alpha=0$ and
transverse Ising model with $\alpha=1$ may now be easily obtained.
%$$ \Gamma_k(t) = \Theta(t -\frac{{\tau}_q}{2}),  For $\alpha=0$,
%\begin{eqnarray}
%~~~~~~~~~~~~~~~~~~~~~~~~~~~~~~~~~~~~~~~~\rm{for}~~~~~~~ %\\
\begin{eqnarray}
 \rm{For}~ \alpha = 0, ~~~~\Gamma_k(t)& = & 2 \pi \Theta(|t| -{{\tau}_q})  \nonumber \\
 \rm{and}~ \rm{for}~ \alpha = 1,~~
 %\begin{equation}
%\begin{equation}
 \Gamma_k(t)& =& \pi \left( 1-\frac{\cos k%\frac{2\pi k}{N}
 +\frac{t}{\tau_q}}{\sqrt{1+\frac{t^2}{\tau_q^2}+2\frac{t}{\tau_q}\cos k}}\right)
 %\end{equation}
\end{eqnarray}
For ${\alpha}=0 $, one of the Rabi frequencies, either (${\Omega_a}$) or (${\Omega_b}$) is zero.
For ${\alpha}= 1$, where both of Rabi frequencies are non zero. 
%  ~~~~\rm{for}~~~~~~\gamma=1 %\frac{2\pi k}{N} .
%\end{eqnarray}
%(t(<0)<<\tau_q
%)%\approx 0~~~~ \rm{or}~~~~ 2\pi N
%\end{equation}
Finally, at $t=0$, when the magnetic field is gradually turned off,
the situation is a bit different.
The configuration of the final state will depend on the number of kinks generated
in the system due to phase transition at or near  $t=-{\tau_q}$ and
as such it will depend on the quench time $\tau_q$ \cite{zur}.
The number of kinks is the number of quasi-particles excited at $B=0$ (
in our parameter space of micro-cavities, ${{\delta}_1}/2 = \beta$, and the
quasiparticle excitation is the excitation of polariton) and is given by
% \begin{equation}
$ {\cal{N}}={\sum_k }p_k $
% \end{equation}
where $p_k$ is the excitation probability (for the slow transition)
and is given by the Landau Zener formula \cite{zener}
%\begin{equation}
$  p_k \approx \exp{ (-2\pi \tau_q k^2)} $.
%\end{equation}
As different pairs of quasi-particles ($k,-k$) evolve independently
for large values of $\tau_q$, it is likely that only one pair of quasi-particles
with momenta $(k_0,-k_0)$ will be excited. Where $k_0(=\frac{\pi}{N})$ corresponds
to the minimum value of the energy $\Lambda_k$.
Thus the condition for adiabatic transition in a finite chain is given by
%\begin{equation}
$ \tau_q >> \frac{N^2}{2 \pi^3} $.
%\end{equation}
Hence, well in the adiabatic regime, the
final state at $t=0$  is given by
%\begin{equation}\label{final}
\bea
 |\psi_{final} > & = & |1>_{k_0}|0>_{-k_0} \non \\
& &  \prod_{k,k\neq\pm k_0}
%\frac{\pi}{N}}
 (\cos\frac{\theta_k}{2}|0>_k|0>_{-k} \non \\
& &  -i\exp(2i\phi)\sin\frac{\theta_k}{2}|1>_k|1>_{-k} )
\eea
\begin{figure}
\includegraphics[scale=0.50,angle=0]{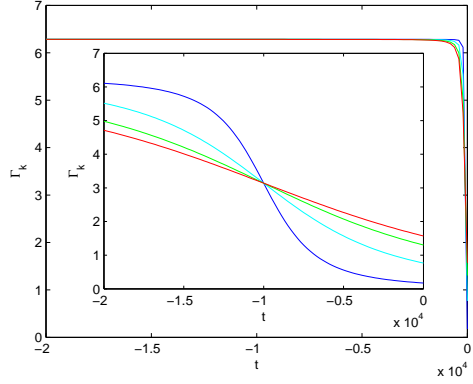}
\caption{Color online, variation of the geometric phase
with time for two different quenching time, $\tau_q =100 , 10,000$ for
four different anisotropy parameters $\alpha= 0.2, 0.5, 0.8, 1.0$ for
the blue, cyan, red, green respectively. Here $ k = \pi/3 $. 
 }
\label{Fig. 2 }
\end{figure}
\begin{figure}
\includegraphics[scale=0.50,angle=0]{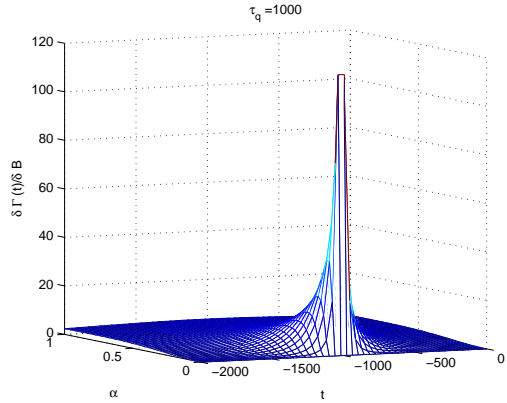}
\caption{Color online, this three dimensional figures shows
the variation of first order derivative of geometric phase with 
time, with the anisotropy exchange
parameter and time. Its shows the existence of XX criticality ($ {\alpha =0}$ ).
 }
\label{Fig. 4 }
\end{figure}
This state is similar to the direct product of only $N-1$ spins oriented along $(2\phi, \theta_k)$
where the state of the spin corresponding to momentum $k_0$ does not contribute to the geometric phase.
 %It can be easily checked that the contribution for the geometric phase due
%to the flipped quasi-particle with momentum mode $(k_0,-k_0)$ is zero.
The total geometric phase of this state is given by
%\begin{equation}
$ \Gamma_{final}(t=0)=\sum_{k, k\neq \pm k_0}\pi (1-\cos\theta_k) $.
For isotropic, $\alpha =0$ and for quantum spin chain system, geometric phase shows a sharp drop
and finally become zero at $t= {\tau_q}$ 
(see the inset of the Fig. 1) for this case optical cavity array system has only
single Rabi frequency oscillation. 
In our study, we set the energy scale $J_1 = 1$, that
leads to the following relations in the Rabi-frequencies, laser frequencies and coupling strengths
in the following way, 
$ \frac{4}{\gamma_2} = ( \frac{{|{\Omega_a}|}^2 {g_b}^2 }{{ {\Delta}_a }^2 }
 +  \frac{{|{\Omega_b}|}^2 {g_a}^2 }{{ {\Delta}_b }^2 } ) $.
We achieve the isotropic $H_{XX}$ Hamiltonian, either ${\Omega}_a $ or ${\Omega}_b $ 
is zero. For ${\Omega}_a =0 $, the reduce analytical expression for  
the magnetic field and $J_1 $ exchange are  
$ B   =  \frac{\delta_1}{2} - \frac{1}{2} [\frac{{|{\Omega_b}|}^2 }{4 {\Delta}_b }
({\Delta}_b - \frac{{|{\Omega_b}|}^2 }{4 {\Delta}_b } -  
 \frac{{|{\Omega_b}|}^2 }{4 ( {\Delta}_a  - {\Delta}_b )} - {\gamma_b} {g_b}^2 
- {\gamma_1} {g_a}^2 + {\gamma_1}^2 \frac{{g_a}^4 }{{\Delta_b}} ] $.
and  
$ {J_1} = \frac{\gamma_2}{4} ( 
  \frac{{|{\Omega_b}|}^2 {g_a}^2 }{{ {\Delta}_b }^2 } ) $ respectively.
For ${\Omega_b} =0$,
the analytical expression for $B$ and $J_1 $ can be obtained from Eq. (7) and Eq. (8)
by putting ${\Omega_b} =0 $. \\
For $\alpha=1 $, the spin chain is in a transverse
magnetic field.
One can achieve this limit by adjusting the detuning field, Rabi frequencies and the
coupling strengths. For this limit, the analytical expression between the Rabi frequency 
and coupling strength is  
$ {\Delta}_b = \frac{ \gamma_2 \Omega_a \Omega_b g_a g_b}{2 \Delta_a } $,
two Rabi frequency oscialltion exist for this case.
The analytical expression for effective magnetic filed and the $J_1 $ coupling can
be obtained by substituting the analytical expression of $\Delta_b $ in 
Eq (7) and Eq. (8). 
The dynamics of the geometric phase shows no sharp drop of the geometric phase through out the quenching period.
The dynamics of the geometric phase under quenching is different for isotropic
quantum spin chain system ($H_{XX}$).
Inset of the Fig. 1 shows the dynamics of geometric phase for $H_{XX}$ model. 
We will also observe  that the first order derivative of the geometric phase also
behave differently for these two limit of our model Hamiltonian.
Therefore it is very clear from this study that the presence of single
Rabi frequency and two Rabi frequency oscillation in the system of cavity
QED changes the dynamics of geometric phase differently.\\
In Fig. 2, we present the study of the geometric phase for different values of anisotropy
exchange interactions ${\alpha} = 0.2, 0.8$ 
and for three
different quenching times ${\tau}_q = 250, 500, 1000$, 
which can be achieved by
manipulating two Rabi frequencies ${\Omega}_a$ and ${\Omega}_b$. 
It reveals from our study that the 
qualitative behavior of the geometric phase is same but for the lower value
of exchange anisotropy decaying rate is higher than for the higher values of
exchange anisotropy. The geometric phase for different anisotropic exchange
parameters crosses at the point $ t= {\tau_q}$ for the same ${\tau}_q$. 
This common crossing point is nothing but the quantum critical
point for the system at $B= 1$ for the same $\tau_q $. 
This has shown explicitly in Fig. 5 of
this study.\\
Apart from that the 
geometric phase for lower value of $\alpha$ also crosses the geometric
phases of other $\tau_q$ but at the time lower than $\tau_q$. Therefore it is
clear from this study due to the smaller values of Rabi frequencies (${\Omega}_a$
or ${\Omega_b}$) decaying rate of geometric phase is sharper compare to
the larger values of Rabi frequencies. The same physics can also be obtained 
by regulating ${\Delta}_a $ and ${\Delta}_b$. It is very clear from the
analytical expression of Eq. 14 and Eq. 15 that the dynamics of geometric
phase depends on the magnetic field (time dependent
quenching field) and the exchange anisotropy, therefore for a fixed $\tau_q $
the common meeting point of the geometric phase occurs at the quenching time. 
\\
In Fig.3, we study the effect of different exchange interactions for $\tau_q =1000$.
We observe that the all geometric phases crosses at the point $t ={\tau}_q $.
It reveals from this study that for a fixed quenching rate, the different exchange
anisotropy (the presence of different Rabi frequencies oscillation)
has no effect to change the common crossing point of the geometric phase. \\
In our study, $1/{\tau_q} $, is the quenching rate for smaller values of
$\tau_q $ is larger and smaller for higher values of $\tau_q$. In Fig. 4
shows the dynamical evalution of geometric phase for higher and lower
quenching rate. It appears from our study that for higher quenching 
rate $(\tau_q = 100 )$ all curves for different anisotropy. For lower
quenching rate is smaller for this case there is no collapse of
the dynamical behaviour of the geometric phase.\\
In Fig. 4, we study the effect of slower and rapid quenching
rate on the dynamical behaviour of geometric phase. It
appears from our study that for higher quenching rate
($\tau_q = 100 $) all curves for different anisotropy merge
into a single line as if the exchange anisotropy,i.e., the 
different values of two Rabi frequency oscillations has no
effect on the dynamical behaviour of the geometric phase.
The dynamical behaviour of the geometric phase for lower
quenching rate ($\tau_q = 10, 000 $) is the same as that
of $\tau_q = 1000 $. The crossing point of the geometric 
phase at $B=1 $ for a fixed $\tau_q $ is the same for 
all quenching rate.\\ 
To study the criticality of the geometric phase, we study the variation of the 
geometric phase
(${\Gamma}_k $) and its derivative with respect to the quench
induced magnetic filed ($B$) i.e. $ ({\frac{d {\Gamma}_k}{d B}})$
with time. We find the non-analytic behavior of the derivative at $
t={\tau}_q$. The analytical expression for the derivative is 
\beq
{\frac{d {\Gamma}_k (t)}{d B}} = \frac{\pi {\alpha}^2 sin^{2} (k) }
{{({( {cos(k) + t/{{\tau}_q})}^2 + {\alpha}^2 sin^{2} (k) })}^{3/2}}
\eeq
Fig. 5, shows the total
variation of $ {\frac{d {\Gamma}_k (t)}{d B}}$. The non-analytical
behavior for $\alpha =0$ at $t= -{\tau_q} $ helps us to predict XX
criticality under the linear quenching process. The analysis with
different values of $\tau_q$ shows that the appearance of XX
criticality is independent of $\tau_q $, i.e.,
independent of fast and slow quenching rates. Therefore, it is
clear from our study that the quantum criticality of the geometric
phase only obtain in the presence of single Rabi frequency oscillation
in the system.\\
{\bf Conclusions:} 
We have studied the quantum criticality of the geometric phase for optical cavity
arrays. We have predicted XX criticality of the geometric phase, which is
independent of the quenching rate. We present the result for the
dynamics of geometric phase for different quenching
rate.  
The presence of two Rabi frequencies oscillation
wash out the quantum critical behavior of the geometric phase. 
\centerline{\bf Acknowledgments}
\vskip .2 true cm
The author would like to thank, Prof. S. Girvin for useful discussions
during the international workshop/school on Dirac Materials at ICTS (December,
2012) and also 
the library of Raman Research Institute (Mr. Manjunath).
Finally, the author would like to thank Dr. P. K. Mukherjee for reading the
manuscript carefully.

\end{document}